\documentclass[twocolumn,showpacs,aps,prl,superscriptaddress,letterpaper]{revtex4}

\usepackage{graphicx}
\usepackage{dcolumn}
\usepackage{amsmath}
\usepackage{epsfig}
\usepackage{multirow}

\long\def\inst#1{\par\nobreak\kern 4pt\nobreak
    {\itshape #1}\par\vskip 10pt plus 3pt minus 3pt}
\RequirePackage{xspace}
\usepackage{relsize}

\def\Bztorhozrhoz {\ensuremath{\Bz \to \rho^0 \rho^0 }\xspace}
\def\Btozz {\ensuremath{\Bz \to \rho^0 \rho^0 }\xspace}

\def\Bztorhoprhom {\ensuremath{\Bz \to \rho^+ \rho^- }\xspace}
\def\Bptorhozrrhop {\ensuremath{\Bp \to \rho^+ \rho^0 }\xspace}

\def\babar{\mbox{\slshape B\kern-0.1em{\smaller A}\kern-0.1em
    B\kern-0.1em{\smaller A\kern-0.2em R}}}
\def\Bbar    {\kern 0.18em\overline{\kern -0.18em B}{}\xspace}
\def\Dbar    {\kern 0.18em\overline{\kern -0.18em D}{}\xspace}
\def\Kbar    {\kern 0.18em\overline{\kern -0.18em K}{}\xspace}
\def\pep2{PEP-II}
\mathchardef\Upsilon="7107
\newcommand{\optbar}[1]{\shortstack{{\tiny (\rule[.4ex]{1em}{.1mm})}
  \\ [-.7ex] $#1$}}
\def\BorBbar    {\kern 0.18em\optbar{\kern -0.18em B}{}\xspace}
\def\DorDbar    {\kern 0.18em\optbar{\kern -0.18em D}{}\xspace}
\def\KorKbar    {\kern 0.18em\optbar{\kern -0.18em K}{}\xspace}

\def\qqbar {\ensuremath{q\overline q}\xspace}
\def\babar{\mbox{\slshape B\kern-0.1em{\smaller A}\kern-0.1em
    B\kern-0.1em{\smaller A\kern-0.2em R}}}
\def\Dbar    {\kern 0.18em\overline{\kern -0.18em D}{}\xspace}
\def\B       {\ensuremath{B}\xspace}
\def\Bbar    {\kern 0.18em\overline{\kern -0.18em B}{}\xspace}

\def\BB      {\ensuremath{B\Bbar}\xspace}
\def\Bz      {\ensuremath{B^0}\xspace}
\def\Bzb     {\ensuremath{\Bbar^0}\xspace}
\def\BzBzb   {\ensuremath{\Bz {\kern -0.16em \Bzb}}\xspace}
\def\Bu      {\ensuremath{B^+}\xspace}
\def\Bub     {\ensuremath{B^-}\xspace}
\def\Bp      {\ensuremath{\Bu}\xspace}

\def\BpBm    {\ensuremath{\Bu {\kern -0.16em \Bub}}\xspace}

\def\CP                {\ensuremath{C\!P}\xspace}
\def\pep2{PEP-II}
\mathchardef\Upsilon="7107
\def\Y#1S{\ensuremath{\Upsilon{(#1S)}}\xspace}

\def\FourS {\Y4S}

\def\BR         {{\ensuremath{\cal B}\xspace}}
\def\Bztorhozrhoz {\ensuremath{\,\Bz \to \rho^0\rho^0}\xspace}
\def\Bztorhoprhom {\ensuremath{\,\Bz \to \rho^+\rho^-}\xspace}

\def\Bztorhozfz {\ensuremath{\,\Bz \to \rho^0 f_0(980)}\xspace}
\def\Bztofzfz   {\ensuremath{\,\Bz \to f_0(980) f_0(980)}\xspace}

\def\DeltaE {\ensuremath{\Delta E}\xspace}
\def\mes{\ensuremath{m_{\mathrm{ES}}}\xspace}

\newcommand{\tev}{\ensuremath{\mathrm{\,Te\kern -0.1em V}}\xspace}
\newcommand{\gev}{\ensuremath{\mathrm{\,Ge\kern -0.1em V}}\xspace}
\newcommand{\mev}{\ensuremath{\mathrm{\,Me\kern -0.1em V}}\xspace}
\newcommand{\kev}{\ensuremath{\mathrm{\,ke\kern -0.1em V}}\xspace}
\newcommand{\ev}{\ensuremath{\mathrm{\,e\kern -0.1em V}}\xspace}
\newcommand{\gevc}{\ensuremath{{\mathrm{\,Ge\kern -0.1em V\!/}c}}\xspace}
\newcommand{\mevc}{\ensuremath{{\mathrm{\,Me\kern -0.1em V\!/}c}}\xspace}
\newcommand{\gevcc}{\ensuremath{{\mathrm{\,Ge\kern -0.1em V\!/}c^2}}\xspace}
\newcommand{\mevcc}{\ensuremath{{\mathrm{\,Me\kern -0.1em V\!/}c^2}}\xspace}

\newcommand{\jprlBase}       {Phys.\ Rev.\ Lett.\xspace}
\newcommand{\jprl}      [1]  {\jprlBase\ {\bf #1}}

\def\qqbar {\ensuremath{q\overline q}\xspace}
\def\u     {\ensuremath{u}\xspace}

\def\d     {\ensuremath{d}\xspace}

\def\b     {\ensuremath{b}\xspace}

\def\pip   {\ensuremath{\pi^+}\xspace}
\def\pim   {\ensuremath{\pi^-}\xspace}

\def\B       {\ensuremath{B}\xspace}
\def\mes        {\mbox{$m_{\rm ES}$}\xspace}
\def\DeltaE     {\mbox{$\Delta E$}\xspace}

\def\epem       {\ensuremath{e^+e^-}\xspace}

\newcommand{\eg}{{\em e.g.}}

\newcommand{\BABARPubYear}     {06}
\newcommand{\BABARPubNumber}  {072}

\newcommand{\SLACPubNumber} {12259}

\begin{document}

\hbox to \hsize{
\vtop{\begin{flushleft}
\end{flushleft}}

\hfil
\vtop{\begin{flushright}
\babar-PUB-\BABARPubYear/\BABARPubNumber\\
SLAC-PUB-\SLACPubNumber \\
hep-ex/0612021
\end{flushright}
}
}
\vspace{0.5\baselineskip}

\title{
\large \bfseries \boldmath
Evidence for \Btozz Decay and Implications for the CKM Angle $\alpha$
}

%
\author{B.~Aubert}
\author{M.~Bona}
\author{D.~Boutigny}
\author{Y.~Karyotakis}
\author{J.~P.~Lees}
\author{V.~Poireau}
\author{X.~Prudent}
\author{V.~Tisserand}
\author{A.~Zghiche}
\affiliation{Laboratoire de Physique des Particules, IN2P3/CNRS et Universit\'e de Savoie, F-74941 Annecy-Le-Vieux, France }
\author{E.~Grauges}
\affiliation{Universitat de Barcelona, Facultat de Fisica, Departament ECM, E-08028 Barcelona, Spain }
% ...
\collaboration{The \babar\ Collaboration}
\noaffiliation


\date{December 11, 2006}


\begin{abstract}
We search for the decays \Btozz, \Bztorhozfz, and \Bztofzfz\
in a sample of about 384 million
$\Upsilon (4S)\rightarrow B\kern 0.18em\overline{\kern -0.18em B}$
decays collected with the $\babar$ detector at the
PEP-II asymmetric-energy $e^+e^-$ collider at SLAC.
We find evidence for \Btozz\ with $3.5\sigma$ significance and
measure the branching fraction
$\BR = (1.07\pm 0.33\pm 0.19)\times 10^{-6}$
and longitudinal polarization fraction
$f_L = 0.87\pm 0.13\pm 0.04$, where the first uncertainty is
statistical, and the second is systematic.
The uncertainty on the CKM unitarity angle
$\alpha$ due to penguin contributions in $B\to\rho\rho$
decays is $18^\circ$ at the $1\sigma$ level.
We also set upper limits on the \Bztorhozfz\ and \Bztofzfz\ 
decay rates. 
\end{abstract}


\maketitle

Measurements of \CP-violating asymmetries in the \BzBzb system 
test the flavor structure of the standard model by 
over-constraining the Cabibbo-Kobayashi-Maskawa (CKM)
quark-mixing matrix~\cite{CabibboKobayashi}.
The time-dependent \CP asymmetry in the decays of \Bz\ or \Bzb\ mesons
to a \CP eigenstate dominated by the tree-level amplitude
$\b \to \u{\bar\u}\d$
measures $\sin 2\alpha_\mathrm{eff}$, where
$\alpha_\mathrm{eff}$ differs from the CKM unitarity
triangle angle $\alpha\equiv
\arg\left[-V_{td}^{}V_{tb}^{*}/V_{ud}^{}V_{ub}^{*}\right]$ by a
quantity  $\Delta\alpha$ accounting for the contributions from 
loop (penguin) amplitudes.
The value of $\Delta\alpha$ can be extracted from an analysis 
of the branching fractions of the $B$ decays into the full 
set of isospin-related channels~\cite{gronau90}.

Branching fractions and time-dependent \CP
asymmetries in $B\to\pi\pi$, $\rho\pi$, and $\rho\rho$
have already provided information on $\alpha$.
Since the tree contribution to the $B^0\to\rho^0\rho^0$
decay is color-suppressed,
the decay rate is sensitive to the penguin amplitude.
The $\Bz\to\rho^0\rho^0$ decay has a much smaller branching 
fraction than $\Bz\to\rho^{+}\rho^{-}$ and $B^{+}\to\rho^{+}\rho^0$
channels~\cite{vvbabar,rho0rhopbelle,rho0rhop2,rhoprhomBF,rhoprhomCS,rhoprhombelle},
and therefore a stringent limit on
$\Delta\alpha$ can be set~\cite{gronau90, rhoprhomBF, falketal}.
This makes the $\rho\rho$ system particularly effective for
measuring~$\alpha$.

In $B\to\rho\rho$ decays the final state is
a superposition of \CP-odd and \CP-even states. 
An isospin-triangle relation~\cite{gronau90} holds for each
of the three helicity amplitudes, which can be separated through
an angular analysis. 
The helicity angles $\theta_1$ and $\theta_2$ are defined 
as the angles between the direction of $\pi^+$ and the direction 
of the \B in the rest system of each of the $\rho^0$ candidates.
The resulting angular distribution
${d^2\Gamma / (\Gamma\,d\!\cos \theta_1\,d\!\cos \theta_2)}$ is
\begin{eqnarray}
{9 \over 4} \left \{ {1 \over 4} (1 - f_L)
\sin^2 \theta_1 \sin^2 \theta_2 + f_L \cos^2 \theta_1 \cos^2 \theta_2 \right\},
\label{eq:helicityshort}
\end{eqnarray}
\noindent where $f_L=|A_0|^2/(\Sigma|A_\lambda|^2)$ is the
longitudinal polarization fraction and
$A_{\lambda=-1,0,+1}$ are the helicity amplitudes.

In this paper we present the first evidence for the \Btozz decay, the
measurement of the longitudinal polarization fraction in this decay, 
and updated constraints on the penguin contribution to the 
measurement of the unitarity angle $\alpha$.


These results are based on data collected
with the \babar\ detector~\cite{babar} at the PEP-II asymmetric-energy
$e^+e^-$ collider~\cite{pep2}.
A sample of $383.6\pm 4.2$ million $\BB$ pairs
was recorded at the $\FourS$ resonance with the center-of-mass 
(c.m.) energy $\sqrt{s} = 10.58$~\gev.
Charged-particle momenta and trajectories are measured in a tracking system
consisting of a five-layer double-sided silicon vertex tracker
and a 40-layer drift chamber,
both within a 1.5-T solenoidal magnetic field.
Charged-particle identification is provided by
measurements of the energy loss
in the tracking devices and by a ring-imaging Cherenkov detector.


We select $\B\to M_1M_2\to(\pi^+\pi^-)(\pi^+\pi^-)$
candidates, with $M_{1,2}$ standing for $\rho^0$ or $f_0$ candidate,
from neutral combinations of four charged tracks that
are consistent with originating from a single vertex near
the $e^+e^-$ interaction point. We veto tracks that are positively
identified as kaons or electrons.
The identification of signal $B$ candidates is based
on several kinematic variables. 
The beam-energy-substituted mass,
$\mes = [(s/2 + {\mathbf {p}}_i\cdot {\mathbf{p}}_B)^2/E_i^2-
{\mathbf {p}}_B^2]^{1/2}$,
where the initial $e^+e^-$
four-momentum $(E_i, {\mathbf {p_i}})$ and the \B
momentum ${\mathbf {p_B}}$ are defined in the laboratory frame, is
centered near the \B mass with a resolution of $2.6~\mev$ for signal
candidates.  
The difference $\DeltaE = E_B^{\rm cm} - \sqrt{s}/2$ between the
reconstructed \B energy in the 
c.m. frame and its known value $\sqrt{s}/2$ has a maximum near zero with a
resolution of $20~\mev$ for signal events. Four other kinematic
variables describe two possible
$\pi^+\pi^-$ pairs: invariant masses $m_{1}$, $m_{2}$
and helicity angles $\theta_1,\ \theta_2$. 

The selection requirements for signal candidates are the following:
$5.245 < \mes < 5.290~\gev$, 
$|\DeltaE|<$ 85~\mev,
$550< m_{1,2} < 1050~\mev$,
and $|\cos\theta_{1,2}|<0.98$.
The last requirement removes a region corresponding to low-momentum 
pions with low and more uncertain reconstruction efficiency.
In addition, we veto the copious decays 
$\Bz\to D^{(*)-}\pip\to(h^+\pim\pim)\pip$, 
where $h^+$ refers to a pion or kaon, by requiring the
invariant mass of the three-particle combination
to differ from the $D$-meson mass by more 
than $13.2~\mev$, or $40~\mev$ if one of the particles is consistent with a
kaon hypothesis. 

We reject the dominant  $\epem\to q\bar{q}\ (q=u,d,s,c)$ (continuum)
background by requiring $|\cos\theta_T| < 0.8$, where $\theta_T$
is the angle between the $B$-candidate thrust axis
and that of the remaining tracks and neutral clusters in
the event, calculated in the c.m. frame.
We further suppress continuum background using 
a neural network discriminant $\mathcal{E}$, which combines a number
of topological variables calculated in the c.m. frame.
Among those are the polar angles of the $B$ momentum vector 
and the $B$-candidate thrust axis with respect to the beam axis.
Other discriminating variables 
include the two Legendre moments $L_0$ and $L_2$ of the energy
flow around the $B$-candidate thrust axis~\cite{bigPRD} 
and the sum of the transverse momenta of all particles in the rest
of the event, calculated with respect to the $B$ direction.

After application of all selection criteria,
$N_{\rm cand}=64843$ events are retained.
On average, each selected event has $1.05$ signal candidates, 
while in Monte Carlo~\cite{GEANT}
samples of longitudinally and transversely
polarized $B^0\to\rho^0\rho^0$ decays
we find $1.15$ and $1.03$ candidates, respectively.
When more than one candidate is present in the same event,
the candidate having the best $\chi^2$ consistency
with a single four-pion vertex is selected. 
Simulation shows that 18\% of longitudinally
and 4\% of transversely polarized $\Btozz$
events are misreconstructed with one or more tracks
not originating from the $B^0\to\rho^0\rho^0$ decay.
These are mostly due to combinatorial background from
low-momentum tracks from the other \B meson in the event.

Further background separation is achieved by
the use of multivariate $B$-flavor-tagging
algorithms trained to identify primary leptons, kaons, soft pions,
and high-momentum charged particles
from the other $B$~\cite{babarsin2beta}.
The discrimination power arises from the difference between
the tagging efficiencies for signal and background in seven
tagging categories ($c_{\rm tag}=1..7$).


We use an unbinned extended maximum likelihood fit to extract
the $B^0\to\rho^0\rho^0$ event yield and fraction of longitudinal
polarization $f_L$. We also fit for the event yields of $B^0\to\rho^0f_0$ 
and $B^0\to f_0f_0$ decays, as well as of several background categories.
The likelihood function is
\begin{equation}
{\cal L} = \exp\left(-\sum_{k}^{} n_{k}\right)\,
\prod_{i=1}^{N_{\rm cand}}
\left(\sum_{j}~n_{j}\,
{\cal P}_{j}(\vec{x}_{i})\right),
\label{eq:likel}
\end{equation}
where $n_j$ is the unconstrained number of events for each event type $j$
($B^0\to\rho^0\rho^0$ , $B^0\to\rho^0f_0(980)$, $B^0\to f_0(980)f_0(980)$,
three background components from B decays, 
and continuum), and ${\cal P}_{j}(\vec{x}_{i})$ is the 
probability density function (PDF) of the variables
$\vec{x}_{i}=\{m_{\rm{ES}}, \Delta E, \mathcal{E},
m_1, m_2, \cos\theta_1, \cos\theta_2, c_{\rm tag}\}_i$
for the $i$th event.

We use simulated events to parameterize 
the background  contributions from \B decays.
The charmless modes are grouped into two
classes with similar kinematic and topological properties:
$B^0\to a_1^{\pm}\pi^{\mp}$ and a combination of other 
charmless modes, including 
$B^0\to \rho^0K^{*0}$, $B^+\to\rho^+\rho^0$, $B\to\rho\pi$, 
and $B^0\to\rho^+\rho^-$.
One additional class accounts for the remaining neutral and
charged $B$ decays to charm modes. 
We ignore any other four-pion final states whose 
contributions are expected to be small
in our invariant mass window.

Since the statistical correlations among the variables are found to be small,
we take each ${\cal P}_j$ as the product of the PDFs for the
separate variables. Exceptions are the kinematic correlation between the two
helicity angles in signal, and mass-helicity correlations in
other $B$-decay classes and misreconstructed signal. 

We use double-Gaussian functions to parameterize the
$m_{\rm{ES}}$ and $\Delta E$ PDFs for signal,
and a relativistic Breit-Wigner functions 
for the resonance masses of $\rho^0$
and $f_0(980)$~\cite{f0mass}.
The angular distribution at production for 
$B^0\to\rho^0\rho^0$, $B^0\to \rho^0f_0$, and $B^0\to f_0f_0$
modes (expressed as a function of the longitudinal 
polarization in Eq.~(\ref{eq:helicityshort}) for 
\Btozz) is multiplied by a detector acceptance function 
${\cal G}(\cos\theta_1, \cos\theta_2)$,
determined from Monte Carlo. 
The distributions of misreconstructed signal events
are parameterized with empirical shapes in a way similar
to that used for $B$ background discussed below.
The neural network discriminant ${\cal E}$ 
is described by three asymmetric
Gaussian functions with different parameters for signal
and background distributions.

The PDFs for  non-signal \B decay modes are
generally modeled with empirical analytical distributions.
Several variables have distributions
identical to those for signal, such as $m_{\rm{ES}}$
when all four tracks come from the same $B$, or $\pi^+\pi^-$
invariant mass $m_{1,2}$ when both tracks come from
a $\rho^0$ meson.
Also for some of the modes the two $\pi^+\pi^-$ pairs
can have different mass and helicity distributions, 
\eg\ when only one of the two combinations 
comes from a genuine $\rho^0$ or $f_0$ meson, 
or when one of the two pairs contains a
high-momentum pion (as in $B\to a_1\pi$). In such cases,
we use a four-variable correlated mass-helicity PDF.

The signal and $B$-background PDF parameters are extracted from
simulation. The Monte Carlo parameters for
$m_{\rm{ES}}$, $\Delta E$, and ${\cal E}$ PDFs are adjusted by
comparing data and simulation in control channels with similar
kinematics and topology,
such as $B^0\to D^-\pi^+$ with $D^-\to K^+\pi^-\pi^-$.
The continuum background PDF parameters are left free in the fit. 
Finally, the discrete $B$-flavor tagging PDFs for signal modes are 
obtained in dedicated fits to events with identified exclusive \B
decays. 
The tagging PDFs for inclusive \B backgrounds are determined
by Monte Carlo and their systematic uncertainties are studied in data. 

\begin{figure}[t]
\begin{center}
\setlength{\epsfxsize}{\linewidth}\leavevmode\epsfbox{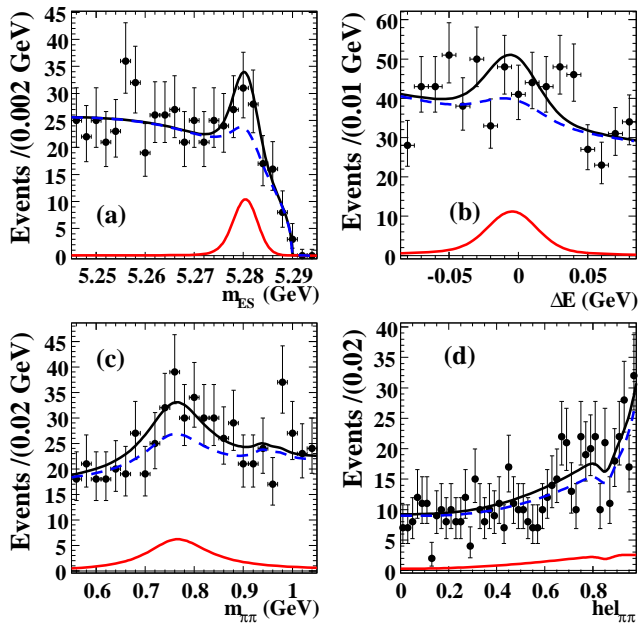}
\end{center}
\caption{
Projections of the multidimensional fit onto
(a) $m_{\rm ES}$, (b) $\Delta E$, (c) di-pion invariant mass
($m_1$ is shown, distribution of $m_2$ is similar),
and (d) cosine of the helicity angle ($\cos\theta_1$ is shown), after
a requirement on the 
signal-to-background probability ratio
with the plotted variable excluded.
This requirement enhances the fraction of
signal events in the sample.
The data points are overlaid by the solid black line,
which corresponds the full PDF projection.
The individual $B^0\to\rho^0\rho^0$ PDF component
is also shown with a solid red line. The sum of all other PDFs
(including $B^0\to\rho^0f_0$ and $B^0\to f_0f_0$ components) is shown
as the dashed blue line. 
The $D$-meson veto causes the acceptance dip seen in (d).
}
\label{fig:projections}
\end{figure}


Table~\ref{tab:results} shows the results of the fit.
The $\Bztorhozrhoz$ decay is observed with a significance of $3.5\sigma$,
as determined by 
the quantity $\sqrt{-2\log(\mathcal{L}_0/\mathcal{L}_{\max})}$, where 
$\mathcal{L}_{\max}$ is the maximum likelihood value, and 
$\mathcal{L}_0$ is the likelihood for a fit with the signal
contribution set to zero. It corresponds to a probability of
background fluctuation 
to the observed signal yield of $2\times10^{-4}$, including systematic
uncertainties, which are assumed to be Gaussian-distributed. We do not
observe significant event yields for $\Bztorhozfz$ and $\Bztofzfz$
decays. Background yields are found to be consistent with
expectations. 
In Fig.~\ref{fig:projections} we show the projections of the fit results
onto $m_{\rm ES}$, $\DeltaE$, $m_1$, and $\cos\theta_1$ variables.

\begin{table}[ht]
\caption{
Summary of results: event yields ($n$);
fraction of longitudinal polarization ($f_L$);
selection efficiency (Eff) corresponding to measured polarization;
branching fraction (${\cal B}_{\rm sig}$);
branching fraction upper limit (UL) at 90\% CL;
and significance including systematic uncertainties.
The systematic errors are quoted last. 
We also show the background event yields for $a_1\pi$, $\qqbar$,
charmless, and other $\BB$ components (statistical uncertainties only). 
}
\vspace{0.2cm}
\begin{tabular}{lc}
\hline\hline
                              &   \vspace*{-0.35cm} \\
Quantity                      &  Value             \\
                              &   \vspace*{-0.35cm} \\
\hline
                              &   \vspace*{-0.3cm} \\
$n$($B^0\to\rho^0\rho^0$)
                              &   $100\pm 32\pm 17$ \\
                              &   \vspace*{-0.35cm} \\
$f_L$ 
                              &   $0.87\pm 0.13\pm 0.04$ \\
                              &   \vspace*{-0.35cm} \\
Eff (\%)                      &   $24.2\pm 1.0$       \\
                              &   \vspace*{-0.35cm} \\
${\cal B}_{\rm sig}$ $(\times 10^{-6})$  &   $1.07\pm 0.33\pm 0.19$ \\
                              &    \vspace*{-0.35cm} \\
Significance, stat. only ($\sigma$)       &    $3.7$   \\
Significance, syst. included ($\sigma$)       &    $3.5$ \\
                              &    \vspace*{-0.35cm} \\
\hline
                              &    \vspace*{-0.3cm} \\
$n$($B^0\to \rho^0f_0$)
                              &    $20\pm 21\ ^{+7}_{-10}$ \\
                              &    \vspace*{-0.35cm} \\
Eff (\%)                      &    $26.1\pm 1.0$       \\
                              &    \vspace*{-0.35cm} \\
${\cal B}_{\rm sig}\times{\cal B}(f_0\to\pi^+\pi^-)$ $(\times 10^{-6})$
                              &    $0.19\pm 0.21\ ^{+0.07}_{-0.10}$ \\
                              &    \vspace*{-0.35cm} \\
UL$\times{\cal B}(f_0\to\pi^+\pi^-)~(\times 10^{-6})$         &   $0.53$  \\
                              &    \vspace*{-0.35cm} \\
\hline
                              &    \vspace*{-0.3cm} \\
$n$($B^0\to\ f_0 f_0$)
                              &   $-3\pm 9\pm 5$ \\
                              &   \vspace*{-0.35cm} \\
Eff (\%)                      &   $28.6\pm 1.1$       \\
                              &   \vspace*{-0.35cm} \\
${\cal B}_{\rm sig}\times{\cal B}^2(f_0\to\pi^+\pi^-)$ $(\times 10^{-6})$
                              &   $-0.03\pm 0.08\pm 0.04$ \\
                              &   \vspace*{-0.35cm} \\
UL$\times{\cal B}^2(f_0\to\pi^+\pi^-)~(\times 10^{-6})$         &   $0.16$  \\
                              &   \vspace*{-0.35cm} \\
\hline
                              &   \vspace*{-0.3cm} \\
$n$($B^0\to a_1^\pm\pi^\mp$)          & $81\pm 25$ \\
                              &   \vspace*{-0.35cm} \\
$n$(${\rm charmless}$)           & $-17^{+107}_{~-96}$   \\
                              &   \vspace*{-0.35cm} \\
$n$(${\BB}$)                     & $3198\pm 224$   \\
                              &   \vspace*{-0.35cm} \\
$n$(${\qqbar}$)                  & $61469\pm 311$  \\
                              &   \vspace*{-0.35cm} \\
\hline\hline
  \end{tabular}
  \label{tab:results}
\end{table}


Dominant systematic uncertainties in the fit originate from 
statistical errors in the PDF parameterizations, due to the limited
number of events in the control samples.
The PDF parameters are varied by their respective uncertainties
to derive the corresponding systematic errors 
($\pm 10$, $^{+6}_{-9}$, $\pm 4$ events for $\rho^0\rho^0$,
$\rho^0f_0$, and $f_0f_0$ 
respectively, and $0.03$ for $f_L$).
We also assign a systematic error of $2$ events for $\rho^0\rho^0$,
3 events for $\rho^0f_0$, and 1 event for $f_0f_0$ ($0.01$ for $f_L$) 
to account for a possible fit bias, evaluated with Monte Carlo experiments.
The above systematic uncertainties do not scale with event yield
and are included in the calculation of the significance of the result.

We estimate the systematic uncertainty due to the interference 
between the $B^0\to\rho^0\rho^0$ and $a_1^{\pm}\pi^{\mp}$ decays using 
simulated samples in which the decay amplitudes for $B^0\to\rho^0\rho^0$
are generated according to this measurement
and those for $B^0\to a_1^{\pm}\pi^{\mp}$ correspond
to a branching fraction of $(33.2\pm4.8)\times 10^{-6}$~\cite{a1pi}.
Their amplitudes are modeled with a Breit-Wigner function
for all $\rho\to\pi\pi$ and $a_1\to\rho\pi$ combinations 
and their relative phase is assumed to be constant across the phase space.
The strong phases and \CP\ content of the interfering state
$a_1^{\pm}\pi^{\mp}$ are varied between zero and a maximum 
value using uniform prior distributions.
We take the RMS variation of the average signal yield
(14 events for the $\rho^0\rho^0$ yield, or $0.03$ for $f_L$) 
as a systematic uncertainty.

Uncertainties in the reconstruction efficiency
arise from track finding (2\%),
particle identification (2\%),
and other selection requirements,
such as vertex probability (2\%),
track multiplicity (1\%),
and thrust angle (1\%).


To constrain the penguin contributions to $B\to\rho\rho$ decays, we
perform an isospin analysis,
by minimizing a $\chi^2$ term that includes the measured quantities
expressed as the lengths of the sides of the isospin triangles.
We use the measured branching fractions and
fractions of longitudinal polarization of the
$\Bptorhozrrhop$~\cite{rho0rhop2}
and $\Bztorhoprhom$~\cite{rhoprhomBF} 
decays, the \CP-violating parameters $S^{+-}_{L}$ and $C^{+-}_{L}$
determined from the time evolution of the longitudinally
polarized $\Bztorhoprhom$ decay~\cite{rhoprhomCS}, 
and the branching fraction and polarization
of $\Bztorhozrhoz$ from this analysis.
We assume uncertainties to be Gaussian and 
neglect $I=1$ isospin contributions, electroweak loop amplitudes, 
non-resonant and isospin-breaking effects.

\begin{figure}[t]
\begin{center}
\setlength{\epsfxsize}{1.0\linewidth}
\leavevmode\epsfbox{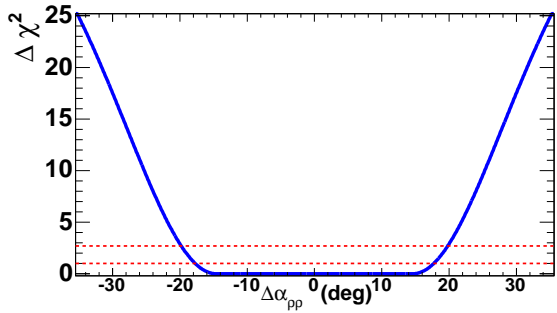}
\caption{$\Delta\chi^2$ as a function of $\Delta\alpha$
obtained from the isospin analysis discussed in the text.
The dashed lines at $\Delta\chi^2 = 1$ and $\Delta\chi^2 = 2.7$
are taken for the $1\sigma$ (68\%) and $1.64\sigma$ (90\%)
interval estimates.}
\label{fig:alphascan}
\end{center}
\end{figure}

With the \Bztorhozrhoz measurement we 
obtain a 68\% (90\%) CL limit on
$|\Delta\alpha|\equiv |\alpha-\alpha_{\rm eff}|<18^\circ$ ($<20^\circ$).
Fig.~\ref{fig:alphascan} shows 
$\Delta\chi^2$ as a function of $\Delta\alpha$.
The central value of $\alpha$ obtained from the isospin
analysis is the same as $\alpha_{\rm eff}$,
which is constrained by the relation
$\sin(2\alpha_{\rm eff})= S^{+-}_{L}/({1-C^{+-2}_{L}})^{1/2}$
and is measured with the $B^0\to\rho^+\rho^-$
decay~\cite{rhoprhomCS}.

The error due to the penguin contribution becomes
the dominant uncertainty in the measurement of $\alpha$ using
$B\to\rho\rho$ decays. However, once the sample 
of $B^0\to\rho^0\rho^0$ decays becomes more significant, 
time-dependent angular analysis will allow us
to measure the \CP\ parameters $S^{00}_{L}$ and $C^{00}_{L}$,
analogous to $S^{+-}_{L}$ and $C^{+-}_{L}$,
resolving ambiguities inherent to isospin triangle orientations.

In summary, we find evidence for \Btozz\ decay with $3.5\sigma$
significance. We measure the $\Btozz$ branching fraction of
$(1.07\pm 0.33\pm 0.19)\times 10^{-6}$
and determine the longitudinal polarization fraction for these
decays of $f_L = 0.87\pm 0.13\pm 0.04$.
The measurement of this branching fraction combined with that for
$B^+\to\rho^0\rho^+$ and $B^0\to\rho^+\rho^-$ decays provides
a constraint on the penguin uncertainty
in the determination of the CKM unitarity angle $\alpha$.
 These results supersede our previous measurements~\cite{vvbabar}. 
We find no significant evidence for the decays $B^0\to \rho^0f_0$
and $B^0\to f_0f_0$.


We are grateful for the excellent luminosity and machine conditions
provided by our \pep2\ colleagues,
and for the substantial dedicated effort from
the computing organizations that support \babar.
The collaborating institutions wish to thank
SLAC for its support and kind hospitality.
This work is supported by
DOE
and NSF (USA),
NSERC (Canada),
IHEP (China),
CEA and
CNRS-IN2P3
(France),
BMBF and DFG
(Germany),
INFN (Italy),
FOM (The Netherlands),
NFR (Norway),
MIST (Russia),
MEC (Spain), and
PPARC (United Kingdom).
Individuals have received support from the
Marie Curie EIF (European Union) and
the A.~P.~Sloan Foundation.



\begin{thebibliography}{99}

\bibitem{CabibboKobayashi}
N. Cabibbo, Phys. Rev. Lett. {\bf 10}, 531 (1963);
M. Kobayashi, T. Maskawa, Prog. Theor. Phys. {\bf 49}, 652 (1973).

\bibitem{gronau90}
M.~Gronau, D.~London, \jprl{65}, 3381 (1990).

\bibitem{footnote}
Charge conjugate $B$ decay modes are implied in this paper.

\bibitem{vvbabar}
\babar\ Collaboration, B.~Aubert {\it et al.},
\jprl{91}, 171802 (2003);
\jprl{94}, 131801 (2005).

\bibitem{rho0rhopbelle}
Belle Collaboration, J.~Zhang {\it et al.},
\jprl{91}, 221801 (2003).

\bibitem{rho0rhop2}
\label{ref:rho0rhop2}
\babar\ Collaboration, B.~Aubert {\it et al.},
arXiv:hep-ex/0607092 (2006), accepted to \jprl.

\bibitem{rhoprhomBF}
\babar\ Collaboration, B.~Aubert {\it et al.},
Phys. Rev. D {\bf 69}, 031102 (2004);
\jprl{93}, 231801 (2004).

\bibitem{rhoprhomCS}
\babar\ Collaboration, B.~Aubert {\it et al.}, 
\jprl{95}, 041805 (2005).

\bibitem{rhoprhombelle}
\label{ref:rhoprhombelle}
Belle Collaboration, A.~Somov {\it et al.},
\jprl{96}, 171801 (2006).

\bibitem{falketal}
A.F.~Falk {\it et al.}, Phys.\ Rev.\ D {\bf 69}, 011502 (2004).

\bibitem{babar}
\babar\ Collaboration, B.~Aubert {\it et al.},
Nucl. Instrum. Methods Phys. Res.,
Sect. A {\bf 479}, 1 (2002).

\bibitem{pep2}
PEP-II Conceptual Design Report, SLAC-R-418 (1993).

\bibitem{bigPRD}
$\babar$ Collaboration, B.~Aubert {\it et al.},
Phys.\ Rev.\ D {\bf 70}, 032006 (2004).

\bibitem{GEANT}
The \babar\ detector Monte Carlo simulation is based
on GEANT4: S. Agostinelli {\it et al.},
Nucl. Instrum. Methods Phys. Res.,
Sect. A {\bf 506}, 250 (2003).

\bibitem{babarsin2beta}
\babar\ Collaboration, B.~Aubert {\it et al.},
\jprl{89}, 201802 (2002).

\bibitem{f0mass}
E791 Collaboration, E. M. Aitala {\it et al.},
Phys. Rev. Lett. {\bf 86}, 765 (2001).

\bibitem{a1pi}
\babar\ Collaboration, B.~Aubert {\it et al.},
\jprl{97}, 051802 (2006).

\end{thebibliography}
\end{document}